\documentclass{article}

\usepackage{arxiv}

\usepackage[utf8]{inputenc} 
\usepackage[T1]{fontenc}    
\usepackage{hyperref}       
\usepackage{url}            
\usepackage{booktabs}       
\usepackage{amsfonts}       
\usepackage{nicefrac}       
\usepackage{microtype}      
\usepackage{lipsum}
\usepackage{graphicx}
\graphicspath{ {./images/} }
\usepackage{amsmath,amssymb,amsfonts}
\usepackage{algorithm2e}
\usepackage{algorithmic}
\usepackage{graphicx}
\usepackage{textcomp}
\usepackage{xcolor}
\usepackage{booktabs}
\hypersetup{colorlinks,
      linkcolor=blue,
      citecolor=blue,
      urlcolor=gray}

\title{A Flexible Recommendation System for Individuals and Groups}
\newtheorem{definition}{Definition}[section]

\author{
 Yacine Mokhtari \\
  IMT Atlantique - Lab-STICC, UMR CNRS 6285\\29238 Brest, France \\
  Intescia  Group\\92130 Issy-les-Moulineaux, France \\
  \texttt{yacine.mokhtari@imt-atlantique.fr} \\
   \And
 Grégory Smits \\
  IMT Atlantique - Lab-STICC, UMR CNRS 6285\\29238 Brest, France \\
  \texttt{gregory.smits@imt-atlantique.fr} \\
}

\begin{document}
\maketitle
\begin{abstract}
Group recommender systems typically rely on either aggregating individual preferences or treating groups as distinct meta-users. However, these methods often suffer from static aggregation strategies or data sparsity issues within group histories. This paper introduces a novel approach, that relies on a GNN-based architecture to learn a dual representation of each user’s preferences, capturing their behavior as an independent individual from one side and as a member of a collective from the other side.
By performing a differential analysis of these individual and group-oriented preferences, our system then determines the behavioral profile of each user when joining a group. Finally, specific preference aggregation strategies are defined to cope with the behavioral profiles of the users composing a group. Consequently, the system is equally capable of delivering precise recommendations to individuals and to arbitrary groups, effectively unifying the two traditional paradigms of recommendation. Experiments on synthetic data simulating diverse group settings and behaviors confirm the flexibility and relevance of the proposed approach compared to state-of-the-art methods.\footnote{Corresponding author: gregory.smits@imt-atlantique.fr}
\end{abstract}


\label{sec:introduction}
When it comes to recommend items to a group of users, two families of approaches have been considered. On the one hand, Persistent Group Recommenders (PGRs)~\cite{cao2018attentive,sankar2020,wu2023,vinh2019}
consider a group as a meta-user and learn its group-specific representations. On the other hand, Occasional Group Recommenders (OGRs)~\cite{amer2009group,9101842,liu2012}
aggregate the pre-computed representations of group members, without considering prior group histories. Both families of approaches have their limitations. PGRs struggle to recommend relevant items to groups that are either unseen or rarely seen. The performance of OGRs mainly depends on the aggregation strategy that they rely on, and they do not take into account variability in user behavior when joining a group.

Beyond these methodological limitations, an important aspect of group dynamics has remained largely underexplored in group recommender systems (GRSs): the adaptability of individual members~\cite{kerr2004group,masthoff2006pursuit}. Users often adjust their preferences according to the group context, and this behavioral trait must be taken into account to determine which items to recommend to the group. Within a group, not all members' preferences are of equal importance~\cite{quijano2013social}. Some act as leaders~\cite{gan2025large}, imposing their preferences, while others are more willing to adjust theirs to those of the rest of the group. For instance, in a family group looking for movies recommendations, preferences of children generally prevail over those of their parents. As another example, users joining reading groups with experts often discard their individual preferences to follow those of the experts. These scenarios highlight the importance of modeling user adaptability as a key factor in enhancing group-level recommendations.

This paper introduces a novel group recommendation strategy that overcomes the aforementioned limitations by explicitly modeling user behavior when joining groups. By analyzing how user preferences shift when interacting with items alone or in a group and how diverse these items are, we can determine their behavioral profile. A Knowledge Graph(KG)-enhanced Graph Attention Network (GAT)-based architecture is proposed to learn two preference representations for each user, as an individual and as a member of a group. Second, based on the analysis of these two representations, a strategy is proposed to determine the group behavioral profile of each user. And third, a recommendation aggregation method inspired by bipolar functions is defined to prioritize the preferences of less adaptable members, ensuring their satisfaction, while incorporating the preferences of adaptable members to refine and nuance group-level recommendations.

After a positioning in Section~\ref{sec:related_work}, as an hybrid strategy gathering the advantages of both PGRs and OGRs, section~\ref{sec:model} then details the theoretical and technical components of the proposed approach. Section~\ref{sec:experiments} details comparison with OGR and PGR state-of-the-art methods, and shows its superiority, especially in terms of flexibility as it generates relevant recommendations for both individuals, seen groups and unseen or rarely seen groups.

The complete implementation of the approach as well as the baselines and datasets used to assess it are available at the following url: \url{https://gitlab.imt-atlantique.fr/research/druper}.

\section{Related Work}
\label{sec:related_work}
There are two main approaches to GRs. So-called OGRs consider that users' preferences remain stable when joining groups. To determine which items to recommend to an arbitrary group, OGRs aggregate the group members' individual preferences, or the recommendations yielded by each member individually~\cite{baltrunas2010,amer2009group}. OGRs are designed not to leverage the history of past interactions of different group compositions or of users as group members. The key distinguishing feature of OGRs is the type of aggregation function they use, which generally ranges from average to minimum or maximum satisfaction.
Oppositely, PGRs rely on the history of past interactions to learn, for each group composition, a representation of their preferences. One of the key models in this category, AGREE~\cite{cao2018attentive} uses an attention mechanism to weight the influence of each group member, modelling group decision-making. Its extension, SoAGREE~\cite{cao_social-enhanced_2021}, integrates social relationships to enhance recommendation accuracy on real-world datasets. GroupSA~\cite{9101842} brings the use of voting mechanism in addition to the use of a social network. GroupIM~\cite{sankar2020} employs a self-supervised learning process to capture member contributions by minimizing the mutual information between the group and its members. And ConsRec~\cite{wu2023}, which achieves state-of-the-art performance, aims to capture group consensus using hypergraphs and multi-view learning. SGGCF~\cite{li2023} uses a user-centered graph, containing users, items and groups, to refine the representation of the preferences of the group. 

It has been shown, in both individual~\cite{wang_kgat_2019} and group~\cite{deng_knowledge-aware_2021} contexts, that the use of an item knowledge graph(KG) leads to more relevant recommendation. The SOTA approach in KG-based GRSs, on simulated data, is KGAG~\cite{deng_knowledge-aware_2021}. 

OGRs thus fail modeling complex group behavior by assuming that user preferences remain constant when engaging with items as part of a group. Conversely, PGR approaches are limited by the availability of large-scale group interaction data and are unable to generalize to newly formed or rarely seen groups.

\section{Flexible Recommendation System}

\label{sec:model}
\begin{figure*}[htb]
  \centering
  \includegraphics[width=\textwidth]{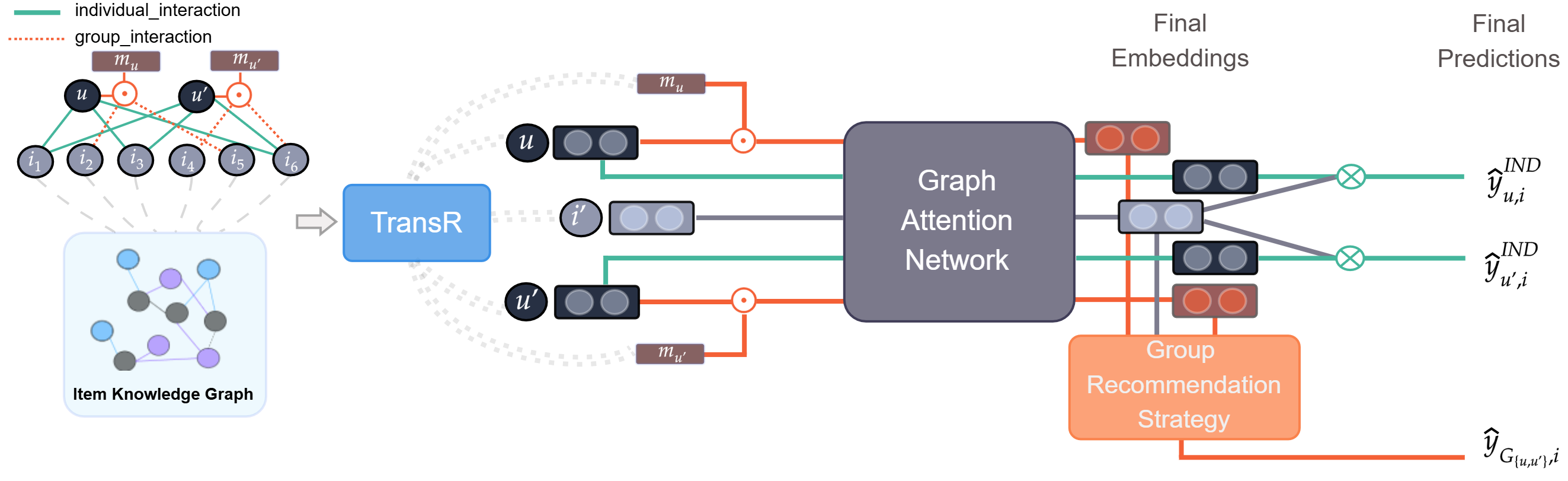}
  \caption{Architecture of the recommendation system.}
  \label{fig:drek_architecture}
\end{figure*}
This section details the components of the proposed approach that make it possible to provide any user and group composition with relevant items.

\subsection{Context and Notations}
\label{model:problem_def}

Let $U$ be set of users, $\mathcal{I}$ be the set of items, and $G$ be the set of groups with known compositions. Items are described as a KG denoted by $\mathcal{G}$. Past interactions between users and items are represented as triplets, $(u, \texttt{individual\_inter}, i)$ means that user $u$ has interacted alone with item $i$ whereas $(u, \texttt{group\_inter}, i)$ means that $u$ has interacted with $i$ as a member of a group. These two types of relations are used to infer two representations of each user's preferences, as an individual and as a member of a group.

\subsection{KG-based GAT Architecture}
\label{subsec:neural_architecture}

\subsubsection{KGE layer} First step is to initiate the latent representation of the entities and relations forming the KG using the TransR model~\cite{lin2015learning}. The model relies on the following scoring function $g$:

\begin{equation}    
    g(h, r, t) = \left\| \mathbf{M}_r \mathbf{e}_h + \mathbf{e}_r - \mathbf{M}_r \mathbf{e}_t \right\|_2^2,
\end{equation}
where $\mathbf{e}_h, \mathbf{e}_t \in \mathbb{R}^d$ are the head and tail entity embeddings, $\mathbf{e}_r \in \mathbb{R}^{d_r}$ is the relation embedding, and $\mathbf{M}_r \in \mathbb{R}^{d_r \times d}$ is the projection matrix for relation $r$.
Notations $h$ and $t$ represent any type of entity in the graph, not only users and items.
A key specificity of the contribution is to learn two representations per user, one as an individual captured in the embedding of the user node, and one as a member of a group. A \textit{modulation vector} $\mathbf{m}_u \in \mathbb{R}^d$ is so attached to each user, representing a non-linear transformation over the user embedding in group setting context. Hence, the scoring function of $u$ is:

\begin{equation}
    g(u, r, i) =
    \begin{cases}
        \|\mathbf{W}_r (\mathbf{e}_u \odot \mathbf{m}_u) + \mathbf{r} - \mathbf{W}_r\mathbf{e}_i\|, &\\
        \text{if } r = \texttt{"group\_inter"} &\\
        \|\mathbf{W}_r \mathbf{e}_u + \mathbf{r} - \mathbf{W}_r\mathbf{e}_i\|,& \text{ otherwise.} \\
    \end{cases}
    \label{transr:scoring_function}
\end{equation}

To learn the embeddings, a contrastive loss is used on a training set $\mathcal{T}_{\text{KGE}}$ composed of positive triplets $(h,r,t) \in \mathcal{G}$ and negative triplets $(h,r,t')$ sampled by corrupting the tail entity such that $(h,r,t') \notin \mathcal{G}$: 

\begin{equation}
\mathcal{L}_{\text{KGE}} = \sum_{(h,r,t,t') \in \mathcal{T}_{\text{KGE}}} -\log \sigma\left( g(h,r,t) - g(h,r,t') \right),
\label{transr:loss}
\end{equation}
where $\sigma(\cdot)$ is the sigmoid function. This loss forces the model to assign higher scores to positive triplets in the KG. This serves as an initial embedding layer and incorporates semantics represented in the domain KG.

\subsubsection{GAT layers} A GAT~\cite{velickovic2017graph} is used to encode structural properties of the KG in the node embeddings. This architecture is particularly suited for learning user-item representations in multi-hop heterogeneous graphs~\cite{wu2022}.

Let us denote by $N_h = \{ t \mid (h,r,t) \in G \}$ the neighborhood of a node $h$. 
At each layer $l$, node embeddings from the neighborhood are aggregated as follows :  

\begin{equation}
\mathbf{e}_{N_h}^{l} = \sum_{(h,r,t) \in N_h} \left(\mathbf{W}_r \mathbf{e}_{t}^{l}\right)^{T} \tanh \left(\mathbf{W}_r \mathbf{e}_{h}^{l} + \mathbf{e}_r\right) \mathbf{e}_{t}^{l} \text{.}
\end{equation}
The following bi-interaction aggregator is applied: 
\begin{align}
f(\mathbf{e}_{h}^{l}, \mathbf{e}_{N_h}^{l}) = &\text{LeakyReLU}\left(\mathbf{W}_{1}^{l} \left(\mathbf{e}_{h}^{l} \mathbf{e}_{N_h}^{l}\right)\right)  \notag    + \\
& \text{LeakyReLU}\left(\mathbf{W}_{2}^{l} \left(\mathbf{e}_{h}^{l} \odot \mathbf{e}_{N_h}^{l}\right)\right) \text{,}
\end{align}

where $\mathbf{W}_{1}^{l}$ and $\mathbf{W}_{2}^{l}$ are trainable weights matrices of layer $l$ to distill useful information for propagation. Thus, the updated embedding at layer $l+1$ is:  

\begin{equation}
\mathbf{e}_h^{l+1} = f\left(\mathbf{e}_h^{l}, \mathbf{e}_{N_h}^{l}\right).
\end{equation}

The initial embedding $\mathbf{e}_h^{(0)}$ is set to $\mathbf{e}_h$ (from Eq. \ref{transr:scoring_function}). After $L$ propagation layers, the final user and item embeddings are obtained:
\begin{equation}
\begin{aligned}
\text{For user } u: & \quad \mathbf{e}_{u}^{\text{\scriptsize IND } (*)} = \mathbf{e}_{u}^{\text{\scriptsize IND } (0)} || \dots || \mathbf{e}_{u}^{\text{\scriptsize IND } (L)}, \\
                    & \quad \mathbf{e}_{u}^{\text{\scriptsize MEM } (*)} = \mathbf{e}_{u}^{\text{\scriptsize MEM } (0)} || \dots || \mathbf{e}_{u}^{\text{\scriptsize MEM } (L)}, \\
\text{For item } i: & \quad \mathbf{e}_{i}^{(*)} = \mathbf{e}_i^{(0)} || \dots || \mathbf{e}_i^{(L)},
\end{aligned}
\end{equation}
with $\mathbf{e}_{u}^{\text{\scriptsize IND } (0)} = \mathbf{e}_{u}^{(0)}$ and $\mathbf{e}_{u}^{\text{\scriptsize MEM } (0)} = \mathbf{e}_{u}^{(0)} \odot \mathbf{m}_{u}$ being the user $u$'s individual and group-member preference respectively at $l=0$.
For each user $u$ and item $i$, Eq.~\ref{eq:predictions} is used to compute an \emph{individual matching score}, denoted $\hat{y}_{u,i}^{\text{\scriptsize IND}}$, and a \emph{group matching score}, denoted $\hat{y}_{u,i}^{\text{\scriptsize MEM}}$. 

\begin{equation}
\hat{y}_{u,i}^{\text{\scriptsize IND}} = \mathbf{e}_i^{(*)^{T}} \mathbf{e}_{u}^{\text{\scriptsize IND } (*)} , \quad \hat{y}_{u,i}^{\text{\scriptsize MEM}} = \mathbf{e}_{i}^{(*)^{T}} \mathbf{e}_{u}^{\text{\scriptsize MEM} (*)}.
\label{eq:predictions}
\end{equation}

These scores are used to optimize the model using Bayesian Personalized Ranking (BPR) loss, known to be suitable for implicit feedback and ranking tasks, which is the case in recommendation systems:

\begin{equation}
\begin{aligned}
    \label{gat:loss_function}
       \mathcal{L}_{\text{CF}} = & \sum_{(u,i,j) \in \mathcal{T}_{\text{\scriptsize IND}}} -\text{ln} \sigma \left(\hat{y}_{u,i}^{\text{\scriptsize IND}} - \hat{y}_{u,j}^{\text{\scriptsize IND}} \right) + \\
       & \sum_{(u,i,j) \in \mathcal{T}_{\text{\scriptsize MEM}}} -\text{ln} \sigma \left(\hat{y}_{u,i}^{\text{\scriptsize MEM}} - \hat{y}_{u,j}^{\text{\scriptsize MEM}} \right) \text{,}
\end{aligned}
\end{equation}
where, $\mathcal{T}_{\text{\scriptsize IND}}$ and $\mathcal{T}_{\text{\scriptsize MEM}}$ represent the training set of individual user interactions and the training set of group-member interactions respectively, made of valid interactions and negative examples sampled randomly.
The joint objective function is written as follows:

\begin{equation}
    \mathcal{L} = \mathcal{L}_{\text{KGE}} + \mathcal{L}_{\text{CF}} + \lambda \|\Theta\|_2^2 \text{ ,}
\end{equation}
where $\Theta = \{ \mathbf{E}, \mathbf{M}, \mathbf{P}, \mathbf{W} \}$ represents the set of model parameters including the entity embeddings ($\mathbf{E}$), the modulation vectors ($\mathbf{M}$), projection matrices $\mathbf{P} = \left\{\mathbf{W}_r, \forall r \in \mathcal{R}\right\}$, and the GAT weight matrices $\mathbf{W} = \left\{\mathbf{W}_{1}^{(l)}, \mathbf{W}_{2}^{(l)}, \forall l \in \left[1..L\right]\right\}$, and $\lambda$ is the L2-regularization coefficient applied to prevent overfitting.

\subsection{Recommending to Groups}
\label{model:group_strategy}
To provide interpretable recommendations, one does not compute the items' relevance scores using an end-to-end learning architecture. Representations of preferences are first learned and then the behavioral profile of each user, to finally determine which items to recommend to arbitrary groups. Categorizing the behavior of each user when joining groups is a crucial step of the process. This profile should indicate how the user adapts their preferences when joining a group. Adaptability is modeled through two complementary aspects: {\it Duality}, which captures the ability of a user to adapt their individual preferences to align with those of other group members, and {\it diversity}, which reflects the variety of items with which a user interacts as a group member.

\begin{definition}[Duality degree]
Let $u$ be a user with two (latent) representations $\mathbf{e}_u^{\text{IND}}, \mathbf{e}_u^{\text{MEM}} \in \mathbb{R}^d$ obtained, respectively, from their interactions with items as an individual and a member of a group respectively. The \emph{duality degree} $\alpha_u \in [0, 1]$ measures the extent to which preferences of user $u$ shift between these two contexts.
\begin{equation}
\label{eq:duality_score}
\alpha_u = \frac{1 - \cos\left( \mathbf{e}_u^{\text{IND}}, \mathbf{e}_u^{\text{MEM}} \right)}{2} = \frac{1 - \frac{ \mathbf{e}_u^{\text{IND}} \cdot \mathbf{e}_u^{\text{MEM}} }{ \| \mathbf{e}_u^{\text{IND}} \| \, \| \mathbf{e}_u^{\text{MEM}} \| }}{2},
\end{equation}
where $\cos$ is the cosine similarity measure.
\end{definition}

\begin{definition}[Diversity degree]
Let $u$ be a user, and let $\mathcal{I}_u^{\text{MEM}}$ denote the set of items $u$ has interacted with in a group context. For each item $i \in \mathcal{I}_u^{\text{MEM}}$, let $\mathbf{e}_i \in \mathbb{R}^d$ be its (latent) representation. The \emph{diversity degree} $\beta_u \in [0, 1]$ captures the degree to which $u$ interacts with diverse items when in group. It is defined as the average pairwise dissimilarity among the item embeddings in $\mathcal{I}_u^{\text{MEM}}$:
\begin{equation}
\label{eq:diversity_score}
\beta_u = \frac{1}{2} - \frac{1}{|\mathcal{I}_u^{\text{MEM}}|(|\mathcal{I}_u^{\text{MEM}}| - 1)} \sum_{\substack{i, j \in \mathcal{I}_u^{\text{MEM}}\\ i \neq j}} \cos\left( \mathbf{e}_i, \mathbf{e}_j \right).
\end{equation}
In case $|\mathcal{I}_u^{\text{MEM}}| < 2$, then $\beta_u$ is set to $0$.
\end{definition}

Based on the duality and diversity scores, and in line with the user categorization described in~\cite{delic2024supporting}, three user profiles are considered:
\begin{algorithm}[b]
\caption{User profiling decision rules}
\label{alg:userprofiling}
\KwData{User $u$, thresholds $\alpha$ and $\beta$}
\KwResult{User profile classification}
\eIf{$dual(u) \geq \alpha$}{
    \eIf{$div(u) \geq \beta$}{
        $u$ is \textbf{AU}\;
    }{
        $u$ is \textbf{DU}\;
    }
}{
    $u$ is \textbf{CU}\;
}
\end{algorithm}

\begin{itemize}
    \item \textbf{Adaptable Users (AUs)}: Users who adjust their preferences depending on the group they join, exhibiting high duality and high diversity scores.
    
    \item \textbf{Non-Adaptable Users (NAUs)}: Users with low adaptability, divided into:
    \begin{itemize}
        \item \textbf{Constant Users (CUs)}: Users whose individual and group preferences remain similar (low duality).
        
        \item \textbf{Dual Users (DUs)}: Users who join groups that differ from their individual preferences (high duality), but whose group interactions are homogeneous (low diversity).
    \end{itemize}
\end{itemize}

Algorithm~\ref{alg:userprofiling} describes the decision-making rules for categorizing user behaviors, where $\alpha$ (resp. $\beta$) is a duality (resp. item diversity) threshold that can be adjusted to cope with the dataset particularities. 

\subsubsection{Aggregation Strategy wrt. Group Configuration}
OGR systems assume that member preferences are commensurable and compensatory, allowing AUs to override the preferences of NAUs. A layered aggregation strategy prioritizes the preferences of NAUs, to determine the set of candidate items to recommend, and then considers the preferences of AUs to re-rank these candidate items. In this sense, the proposed group recommendation strategy acts as a behavioral-driven bipolar aggregation function~\cite{dubois2008introduction}.

Let $g$ be a group of users, $g_a \subseteq g$ its subgroup of AUs, $g_c \subseteq g$ its subgroup of CUs and $g_d \subseteq g $ its subgroup of DUs, such that $g_a \bigcup g_c \bigcup g_d = g$. As detailed in Table~\ref{tab:groupcomposition}, three different group compositions are considered and handled with dedicated recommendation strategies. Let $i$ and $i'$ be two items and $g$ a group of users. $least(g)$ is the least-misery aggregation strategy that ensures the satisfaction of all members of $g$. Formally:
\begin{equation} 
i \preceq i' \text{ if } min_{u \in g}\, \hat{y}_{u,i}^{\text{\scriptsize MEM}} \geq min_{u \in g}\, \hat{y}_{u,i'}^{\text{\scriptsize MEM}},
\end{equation}
where $ i \preceq i'$ means that $i$ is ranked before $i'$.
$avg(g)$ is the average aggregation strategy looking for an overall compensatory satisfaction of $g$s' members. More formally, using the average strategy:
\begin{equation} 
i \preceq i' \text{ if } \frac{1}{|g|} sum_{u \in g} \hat{y}_{u,i}^{\text{\scriptsize MEM}} \geq \frac{1}{|g|} sum_{u \in g} \hat{y}_{u,i'}^{\text{\scriptsize MEM}}.
\end{equation}

The $bipolar(least(g), avg(g'))$ operator uses the preferences of $g$ to build the set $\top^K_{g}$ of candidate items to recommend, and the preferences of $g'$ to rank these candidates. More formally:
\begin{equation}
    \top^K_{g}  = \left\{i \in I \mid\not\exists i' \in I \setminus \top^K_{g} , least(g,i) < least(g),i')\right\}\text{.}
\end{equation}

In the recommendation strategy for group composition $G_1$, preferences of DUs, despite being non-adaptable, are combined with those of AUs to rank candidate items filtered by CUs' preferences. The following justifies the non-prioritization of DUs' preferences when at least one CU is present in the group. Either the group preferences of the DU(s) align with the constant preferences of the CU(s), thus confirming an interest for specific items when in group and justifying the low diversity characterizing profiles of type DU. In this case, the preferences of both the CUs and the DUs are satisfied. Alternatively, the low diversity of such dual users may be due to a poor group context in the training set, leading to an overconsideration of their duality. In this case, aggregating the DUs' group preferences with those of AUs would make sense.
\begin{table}[t]
    \centering
    \caption{Group compositions according to members' profiles}
    \begin{tabular}{ccc}
    \toprule
    Type & Group composition & Aggregation strategy\\
    \midrule
   $G_1$ & $g_c \neq \emptyset$; $g_d \neq \emptyset$;  & $bipolar(least(g_c), avg(g \cup g_c))$\\    
   $G_2$ & $g_c \neq \emptyset$; $g_d = \emptyset$;  & $bipolar(least(g_c), avg(g_a))$\\    
    $G_3$ & $g_c = \emptyset$; $g_d \neq \emptyset$;  & $bipolar(least(g_d), avg(g_a))$\\    
    $G_4$ & $g_c = \emptyset$; $g_d = \emptyset$;  & $avg(g_a)$\\    
    \bottomrule
    \end{tabular}
    \label{tab:groupcomposition}
\end{table}
Unlike compensatory aggregation operators, using a bipolar-based aggregation strategy for group compositions (which include both AUs and NAUs) also facilitates explaining the rationale behind a recommendation; that non-adaptable users are prioritized, while efforts are made to satisfy the rest of the group.

\section{Experiments}
\label{sec:experiments}
This section presents the experiments conducted to determine how does the approach compare to baselines, when recommending to individuals, as well as seen and unseen group compositions?

\subsection{Experimentation context}

\subsubsection*{Datasets}
\begin{table}[t]
    \caption{Statistics of MovieLens-KG}
    \centering
    \begin{tabular}{lcccc}
        \toprule
        & \#\textbf{items} & \#\textbf{entities} & \#\textbf{relations} & \#\textbf{triplets} \\
        \midrule
        \textbf{MovieLensKG} & 1598 & 33029 & 24 & 91631 \\
        \bottomrule
    \end{tabular}
    \label{table:mlkg}
\end{table}

\begin{table}[t]
    \centering
    \caption{Statistics about the generated dataset. {\it Nb}: number of users and groups, {\it NbI}: mean number of user/group interactions, {\it NbA} mean number of group affiliations}
    \begin{tabular}{l ccc cccc}
        \toprule
        & \multicolumn{3}{c}{\textbf{Users}} & \multicolumn{4}{c}{\textbf{Group composition types}} \\
        \cmidrule(lr){2-4} \cmidrule(lr){5-8}
        & \textit{AU} & \textit{CU} & \textit{DU}  & \textit{G1} & \textit{G2} & \textit{G3} & \textit{G4} \\
        \midrule
        \multicolumn{8}{c}{\textbf{$\text{genData}_{large}$}}\\
        \midrule
        \textbf{Nb} & 799 & 208 & 215  & 356 & 307 & 296 & 241 \\
        \textbf{NbI} & 35.11 & 33.06 & 32.91  & 41.82 & 41.68 & 42.24 & 42.17 \\
        \textbf{NbA} & 5.08 & 3.4 & 3.23  & --- & --- & --- & --- \\
        \midrule
        \multicolumn{8}{c}{\textbf{$\text{genData}_{small}$}}\\
        \midrule
        \textbf{Nb} & 784 & 209 & 209  & 368 & 307 & 286 & 239 \\
        \textbf{NbI} & 3.49 & 2.95 & 3.03  & 3.46 & 3.54 & 3.43 & 3.64 \\
        \textbf{NbA} & 5.06 & 3.45 & 3.42  & --- & --- & --- & --- \\
        \midrule
        \multicolumn{8}{c}{\textbf{$\text{genData}_{small\ 2}$}}\\
        \midrule
        \textbf{Nb} & 783 & 215 & 216  & 336 & 315 & 309 & 240 \\
        \textbf{NbI} & 14.87 & 13.22 & 13.56  & 3.60 & 3.45 & 3.50 & 3.45 \\
        \textbf{NbA} & 5.06 & 3.25 & 3.31  & --- & --- & --- & --- \\
        \bottomrule
    \end{tabular}
    \label{table:GIKR}
\end{table}

A major limitation when it comes to evaluate GRSs, especially those leveraging an item KG, is the lack of real datasets. CAMRa2011 and Mafengwo are, to the best of our knowledge, the only two real datasets providing both user-item interactions, group-item interaction and group compositions. However, they do not come with an item-KG. The dataset generator dedicated to group recommendation introduced in~\cite{mokhtari2025} has thus been used on top of the movie KG MovieLens-KG~\footnote{\href{https://recbole.io/dataset\_list.html}{https://recbole.io/dataset\_list.html}} to produce three datasets. MovieLens-KG is the main KG used to compare KG-based RS. The generator has been configured to reproduce the data distribution, user behaviors and group compositions observed in CAMRa2011 and Mafengwo. The former exhibits low diversity and high duality, while the latter exhibits high diversity and low duality. They are thus composed of homogeneous dual users (DUs) for the former and constant users (CUs) for the latter. To have a complete representation of the different user behaviors, adaptive users (AUs) are also generated. 

Statistics about the MovieLens-KG are given in Table~\ref{table:mlkg} and about the generated datasets in Table~\ref{table:GIKR}. It shows that both datasets involve very different group compositions but differs mainly on the number of interactions per group. Groups in $\text{genData}_{small}$ and $\text{genData}_{small\ 2}$ have a poor interaction history with items to challenge PGRs. Users in $\text{genData}_{small\ 2}$ have larger individual histories of interaction then those of $\text{genData}_{small}$. Oppositely, groups in $\text{genData}_{large}$ have large histories of interactions with items to assess the capacity of the compared GRS to generalize group behavior.

For individual interactions, we used an $80/20$ train-test split. For group interactions, $10\%$ of groups are not present in the training subset to challenge GRS on unseen group compositions. The remaining $90\%$ of groups were split $70/30$ for training and testing.

\subsubsection*{Experimental Settings}

Items a user has not interacted with are considered as negative during training. For the evaluation, the ranked list of recommendations produced by ranking all the items except the ones used in the training, for each user and group, are evaluated using Hit Rate (HR) and Normalized Discounted Cumulative Gain (NDCG) at $K=20$.
The following baseline models are considered:
\begin{itemize}
    \item \textbf{KGAT:}~\cite{wang_kgat_2019}~Knowledge Graph Attention Network, an explainable SOTA RS using attention, also combined with Least Misery (LM), Maximum Satisfaction (MS) and AVeraGe (AVG) aggregation strategies.
    \item \textbf{KGAT-PGR:} A variant of KGAT where a preference vector is learned for each user and each group composition.
    \item \textbf{KGAG:}~\cite{deng_knowledge-aware_2021}     Knowledge-Aware Group Representation Learning for Group Recommendation, a recent SOTA model.
\end{itemize}

The model was optimized using the Adam optimizer with a batch size of $1024$. All parameters were initialized using Xavier Initialization, except for modulation vectors, which were initialized as vectors of ones. Hyper-parameters were finetuned via grid search over $\{[64, 32, 16],$ $[32, 16, 8],$ $[16, 8, 4]\}$ for the GNN layers, $\{0.1, 0.2, 0.5\}$ for the dropout rate, and $\{0.1,$ $0.05,$ $0.001,$ $0.005,$ $0.0001\}$ for the regularization term. All baselines are also fine-tuned.

\subsection{Performance Evaluation}

\begin{table*}[t]
    \centering
    \caption{Performance comparison to baseline models on diverse generated datasets.}
    \fontsize{8}{9}\selectfont
    \renewcommand{\arraystretch}{1.1}
    \begin{tabular}{llccccccc}
        \toprule
        & & \multicolumn{3}{c}{\textbf{OGR}} 
          & \multicolumn{2}{c}{\textbf{PGR}} 
          & \textbf{Proposed} \\
        \cmidrule(lr){3-5} \cmidrule(lr){6-7} \cmidrule(lr){8-8}
        \textbf{\ } & \textbf{Metric} 
             & \textit{KGAT-lm} & \textit{KGAT-ms} & \textit{KGAT-avg} & \textit{KGAT-PGR} & \textit{KGAG} & \textit{Approach} \\
         \midrule
        \multicolumn{8}{c}{\textbf{$\text{genData}_{small}$}}\\
        \midrule
        \textbf{Inds.} & \textit{HR@20} & 0.044 & 0.044 & 0.044 & 0.041 & 0.051 & \underline{\textbf{\textit{0.052}}} \\
                    & \textit{NDCG@20}    &  0.011 & 0.011 & 0.011 & 0.009 & \textbf{0.013} & \underline{\textbf{\textit{0.013}}} \\
        \textbf{Seen} & \textit{HR@20} & 0.055 & 0.087 & 0.066 & 0.018 & \textbf{0.162} & \underline{\textit{0.132}} \\
        \textbf{Grps.} & \textit{NDCG@20}    & 0.015 & 0.023 & 0.016 & 0.018 & \textbf{0.044} & \underline{\textit{0.025}} \\
        \textbf{Unseen} & \textit{HR@20} & 0.058 & 0.225  & 0.108 & - & - & \underline{\textbf{\textit{0.033}}} \\
        \textbf{Grps.} & \textit{NDCG@20}    & 0.007 & 0.032 & 0.021 & - & - & \underline{\textbf{\textit{0.166}}} \\
         \midrule
        \multicolumn{8}{c}{\textbf{$\text{genData}_{small \ 2}$}}\\
        \midrule
        \textbf{Inds.} & \textit{HR@20} & 0.327 & 0.327 & 0.327 & 0.329 & 0.356 & \underline{\textbf{\textit{0.393}}} \\
                    & \textit{NDCG@20}    &  0.052 & 0.052 & 0.052 & 0.052 & 0.071 & \underline{\textbf{\textit{0.123}}} \\
        \textbf{Seen} & \textit{HR@20} & 0.148 & 0.211 & 0.171 & 0.320 & \textbf{0.346} & \underline{\textit{0.335}} \\
        \textbf{Grps.} & \textit{NDCG@20}    & 0.036 & 0.056 & 0.043& 0.089 & \textbf{0.102} & \underline{\textit{0.095}} \\
        \textbf{Unseen} & \textit{HR@20} & 0.158 & 0.350 & 0.400 & - & - & \underline{\textbf{\textit{0.416}}} \\
        \textbf{Grps.} & \textit{NDCG@20}    & 0.019 & 0.047 & 0.053 & - & - & \underline{\textbf{\textit{0.066}}} \\
                    \midrule
        \multicolumn{8}{c}{\textbf{$\text{genData}_{large}$}}\\
        \midrule
        \textbf{Inds.} & \textit{HR@20} & 0.462 & 0.462 & 0.462 & 0.620 & 0.659 & \underline{\textbf{\textit{0.674}}} \\
                    & \textit{NDCG@20}    &  0.061 & 0.061 & 0.061 & 0.103 & 0.121 & \underline{\textbf{\textit{0.123}}} \\
        \textbf{Seen} & \textit{HR@20} & 0.712 & 0.792 & 0.729 & 0.765 & \textbf{0.832} & \underline{\textit{0.785}} \\
        \textbf{Grps.} & \textit{NDCG@20}    & 0.093 & 0.151 & 0.161 & 0.138 & \textbf{0.184} & \underline{\textit{0.178}} \\
        \textbf{Unseen} & \textit{HR@20} & 0.533 & 0.783 & 0.791 & - & - & \underline{\textbf{\textit{0.883}}} \\
        \textbf{Grps.} & \textit{NDCG@20}    & 0.066 & 0.155 & 0.137 & - & - & \underline{\textbf{\textit{0.178}}} \\
        \bottomrule
    \end{tabular}
    \label{tab:comparison_ogr_pgr_drek}
\end{table*}

Table~\ref{tab:comparison_ogr_pgr_drek} reports the performance comparison and various baselines across three settings: individual recommendations, seen group recommendations, and unseen group recommendations. 

\textbf{Individual Recommendations:} Our approach consistently outperforms all baselines across evaluation metrics at top-$20$ values. Its advantage stems from modeling two distinct user representations, which enhances the quality of collaborative signals. In contrast, KGAT-PGR captures only a single user representation, while variants like KGAT-LM and KGAT-ms make the limiting assumption that group and individual interactions are equivalent.
    
\textbf{Seen Group Recommendations:} When a large amount of interactions are available, KGAT-PGR obviously remains slightly better overall, with the ability to learn specific preference representations for each group. The competitive performance we achieved, which aggregates group interactions per user instead of learning a representation per group, confirms the relevance of the user profiling and adaptive aggregation strategies based on group composition.

\textbf{Unseen Group Recommendations:} Unlike PGR models, the proposed approach generalizes well to unseen groups and provides more accurate recommendations and outperforms OGR models. Where PGRs fail to make recommendations because the groups were not previously observed during the training set.

\section{Conclusion \& Perspectives}
\label{sec:conclusion}

A novel approach to group recommendation is introduced that learns two representations of each user's preferences, one when interacting alone with items and a second one as a member of a group. Comparing these representations and the items that yield them makes it possible to determine the adaptability of the user when joining groups. To recommend items to a group of users, it gives priority to the preferences of non-adaptive users but without discarding adaptive group members. Experiments in various group settings confirm the flexibility of the proposition as it suggest relevant items to individuals, known groups (seen during the model training) as well as unseen groups.

Experiments involving real datasets and possibly synthetic knowledge graphs (KGs) are currently being conducted to confirm the relevance, flexibility and performance of this new approach to GRS compared with state-of-the-art methods. These experiments will be completed with a robustness analysis, wrt. settings of the hyper-parameters, and an ablation study of the different components.

\bibliography{references}
\bibliographystyle{acm}

\end{document}